\documentclass[12pt]{article}
\makeatletter
\def\@cite#1#2{({#1\if@tempswa , #2\fi})}
\makeatother

\newcommand\Eq[1]{\hbox{Eq.(\ref{#1})}}

\newcommand\Fig[1]{Fig.$\:$\ref{#1}}
\newcommand\Figure[1]{Figure$\:$\ref{#1}}

\newcommand{\squishlist}{
   \begin{list}{$\bullet$}
    { \setlength{\itemsep}{0pt}      \setlength{\parsep}{0pt}
      \setlength{\topsep}{3pt}       \setlength{\partopsep}{0pt}
      \setlength{\leftmargin}{1.5em} \setlength{\labelwidth}{1em}
      \setlength{\labelsep}{0.5em} } }
\newcommand{\squishend}{
    \end{list}  }

\usepackage{graphicx}
\usepackage[percent]{overpic}
\usepackage{amsmath}
\usepackage{amssymb}
\renewcommand\Re{\operatorname{Re}}

\begin{document}

\ 
\vskip 1em

\centerline {\bf  Design and application of robust rf pulses for}
\centerline{\bf toroid cavity NMR spectroscopy}

\medskip

\centerline {Thomas E. Skinner $^a$, Michael Braun $^b$, Klaus Woelk$^c$,}
\centerline {Naum I. Gershenzon $^a$ and Steffen J. Glaser $^b$}

\medskip

$^a$ Physics Department, Wright State University, Dayton, OH, 45435 USA\\
$^b$ Department of Chemistry, Technische Universit\"at M\"unchen, 85747 Garching, Germany\\
$^c$ Department of Chemistry, Missouri University of Science and Technology, Rolla, MO 65409 USA\\





\begin{abstract}
\noindent We present robust radio frequency (rf) pulses that tolerate a factor of six inhomogeneity in the B$_1$ field, significantly enhancing the potential of toroid cavity resonators for NMR spectroscopic applications.  Both point-to-point (PP) and unitary rotation (UR) pulses were optimized for excitation, inversion, and refocusing using the gradient ascent pulse engineering (GRAPE) algorithm based on optimal control theory.  In addition, the optimized parameterization (OP) algorithm applied to the adiabatic BIR4 UR pulse scheme enabled ultra-short (50$\mu$s) pulses with acceptable performance compared to standard implementations.  OP also discovered a new class of non-adiabatic pulse shapes with improved performance within the BIR4 framework.  However, none of the OP-BIR4 pulses are competitive with the more generally optimized UR pulses.  The advantages of the new pulses are demonstrated in simulations and experiments.  In particular, the DQF COSY result presented here represents the first implementation of 2D NMR spectroscopy using a toroid probe.
\end{abstract}

\medskip

{\bf keywords}: optimal control theory, GRAPE algorithm, OP algorithm, toroid NMR, composite pulses, shaped pulses, B$_1$ inhomogeneity
\vskip 4em

\vfill \eject
\section{Introduction}
Practical NMR applications require pulses that provide robust performance with respect to experimental limitations, such as resonance offset effects and rf inhomogeneity.  A simple rectangular pulse delivered with a perfectly homogeneous rf-amplitude of 20 kHz (pulse length 12.5 $\mu$s) transforms $M_z$ to $M_x$ with a fidelity of 99\% over an offset range of only $\pm 2.8$ kHz. Miscalibration or inhomogeneity of the $B_1$ field exceeding $\pm\,9$\% reduces $M_x$ on resonance below the desired fidelity.

Increased tolerance to offset and/or rf inhomogeneity can be achieved using
composite \cite{composite1,composite2} and shaped \cite{shaped1} pulses. More recently, efficient pulse design using optimal control theory \cite{Bryson-Ho} 
has made it possible to establish physical limits to pulse performance \cite{BEBOP1,BEBOP2,BEBOP3,Limits, pattern, robust_exc, construction, RC-BEBOP, ICEBERG, Limits2}.  Short ($100\ \mu$s), broadband, PP excitation pulses (98\% fidelity) have been optimized that achieve bandwidth to peak rf ratios of 2 
with a $\pm\,10$\% tolerance to rf inhomogeneity applicable to modern high resolution NMR probes  \cite{Limits}.
In an extreme case study of dual compensation, a 1~ms pulse was found that provides excellent broadband excitation over an offset range of 50~kHz for miscalibration of the B$_1$ field anywhere in the range 10--20~kHz \cite{robust_exc}.  Much larger spatial variations of the B$_1$ field limit appplications involving, e.g., surface coils, ex situ NMR, and high field imaging.

Here we focus on toroid cavity probes \cite{Glass:1983, Woelk:1994, Toroid_Review}, which can be built to tolerate high rf amplitude, high pressure, and high temperature, but at the price of large rf field inhomogeneity. 
A toroid cavity detector is a unique NMR resonator with a wide range of potential applications for in situ reaction studies at high-pressure and/or high-temperature \cite{Woelk:2002}.  Whereas the defined $B_1$ field gradient of the coil is an advantage for some applications (for example, rotating frame NMR imaging on the micrometer scale \cite{Woelk:1993}), it is a serious problem in spectroscopic applications, severely limiting the potential of toroid NMR.
For the toroid probe designs considered here, the ratio between the minimum and maximum B$_1$ field in the sample volume is a factor of six.  Other detector geometries produce even larger inhomogeneities \cite{Toroid_Review}.

We present efficient point-to-point (PP) and unitary rotation (UR) pulses optimized for toroid probe applications using the GRAPE algorithm \cite{GRAPE}. PP pulses rotate one specified initial state about a fixed axis to a desired target state, whereas UR pulses transform any orientation of the initial magnetization about the same axis.  While composite PP pulses developed specifically for toroid detectors exist, \cite{Woelk:1995}, no UR pulses, which are crucial for multi-dimensional NMR, have been developed previously for these probes.  The performance of individual toroid pulses is characterized both theoretically and experimentally in section 3.
We demonstrate the performance and relevance of the new pulses for toroid probe NMR in a DQF COSY experiment, where significant gains in signal amplitude are found compared to experiments using conventional pulses. 

\vskip 1 em
\section{Toroid probes}

The first use of toroid coils in NMR spectroscopy was reported in
1983 \cite{Glass:1983,Glass:1983a}.  Toroid rf coils and toroid cavities provide excellent signal-to-noise ratio
compared to conventional Helmholtz or saddle coils, and  very high rf amplitudes on the order of  100 kHz can be reached. 
In toroid cavity autoclaves  \cite{Woelk:2002,tcapatent}, pressures of up to
300 bar and temperatures of up to
$250\ ^{\circ}\mathrm{C}$ are attainable, making it possible, e.g., to investigate in situ reaction dynamics of homogeneously catalyzed reactions, such as the cobalt-catalyzed hydroformylation process \cite{Klingler:1992,rathke:1991,rathke1991a,Niessen:2002}.

In a typical toroid probe with cylindrical symmetry, the sample is
located between a minimum radius $r_{min}$ (given by the radius of the central conductor) and a maximum radius $r_{max}$.
The $B_1$ field for a toroidal geometry varies inversely with radial distance $r$ as detailed more fully in \cite{Toroid_Review, Woelk:2000, Momot:2000}. 
Hence, the corresponding rf amplitude, $\nu_{rf}(r)=\gamma B_1(r)/(2 \pi)$, 
can be expressed relative to its smallest value at $r_{max}$ as $\nu_{rf}(r)  \propto \nu_{rf}(r_{max}) / r$ to obtain
	\begin{equation}
\nu_{rf}(r) = \frac{r_{max}}{r}\, \nu_{rf}(r_{max}).
\label{nu_rf(r)}
	\end{equation}
The time dependence of the rf pulse can then be written simply in terms of an amplitude modulation function $0 \le a(t) \le 1$ as
	\begin{equation}
\nu_{rf}(r,t) = a(t)\,\nu_{rf}(r)
\label{rf(r,t)}
	\end{equation}
applied to the rf spatial profile at any position $r$ in the toroid.

For the toroid probe used here, $r_{min}=1$~mm and $r_{max}=6$~mm.  A conventional rectangular pulse thus rotates spins near the central conductor by a flip angle that is six times larger than the flip angle experienced by spins near the outer wall of the toroid resonator.  
This large and well defined rf inhomogeneity of toroid probes can be exploited, e.g., in spatially resolved diffusion measurements and imaging \cite{Woelk:1993,Trautner:2002a,Woelk:1996,Trautner:2002b}. However, the large rf inhomogeneity has limited spectroscopic applications of toroid probes to relatively simple experiments.  In the following, we remove this limitation by developing pulses with the necessary high tolerance to rf inhomogeneity.

\vskip 1 em
\section{Pulse optimizations and applications}

The GRAPE algorithm for pulse optimization is discussed in detail in the cited references on optimal control.  A quality factor, $\Phi$, for pulse performance is defined which, in turn, provides an efficiently calculated gradient for iterative improvement of pulse performance.  Most generally, the quality factor is a quantitative comparison between the state of the system and some desired target state.  The gradient therefore also depends on the system state---in the present case, the magnetization $\mathbf{M}$.  Modifications in the basic algorithm that are required for the large and well-defined rf spatial inhomogeneity in toroid probes requires some elaboration.

\subsection{Simulating pulse performance in a toroid}

For a general rf pulse, each combination of offset, $\nu_{off}$, and rf amplitude, $\nu_{rf}(r)$, produces a potentially different transformation of the initial magnetization.  The goal of pulse optimization is to find a particular rf pulse that produces the same transformation for all the scalings of the rf amplitude due to spatial inhomogeneity and all the desired offsets.  A gradient giving the proportional adjustment to make in the rf pulse components to improve performance at each $\nu_{off}$ and $\nu_{rf}(r)$ can be efficiently calculated for point-to-point (PP) pulses \cite{BEBOP1, GRAPE} and for unitary rotation (UR) pulses \cite{GRAPE}. The total gradient is obtained by averaging these constituent gradients over offset and rf inhomogeneity.  

For the small volumes and relatively small deviations from homogeneity seen in standard NMR probes, giving equal weight to different possible spatial values of the rf is sufficient to provide accurate simulations of pulse performance relative to experiment.  In a toroid probe, the effect of the large and well-defined spatial inhomogeneity on both the spatial dependence of the transformed magnetization and the spatially dependent detection sensitivity must be considered.  

The signal from the toroid cavity depends on the total contribution from spins at each radius.  Signal is proportional to the detection sensitivity per spin times the number of spins within a cylindrical sample slice of height $h$, inner radius $r$, and outer radius $r+ \delta r$.  The slice volume $\delta V=2\pi r h\,\delta r$ (and, thus, the corresponding number of spins in the slice) increases linearly with $r$. By the principle of reciprocity \cite{reciprocity}, the detection sensitivity is proportional to $\nu_{rf}$ and hence proportional to $1/r$ (c.f. Eqs. 1 and 2).
Therefore, the signal from a cylindrical slice volume $\delta V$ is {\it independent} of $r$, and we can define an effective magnetization vector representing the sample in a toroid probe as
	\begin{equation}
{\bf M}^{eff}(t)= \frac{1}{r_{max}-r_{min}} \int_{r_{\it min}}^{r_{\it max}} {\bf M}(r,t)\,{\rm d}r
\label{MeffInt(r)}
	\end{equation}
with equal weighting at each radius.  The $(x,y)$ components of the detected signal
are proportional to the $(x,y)$ components of ${\bf M}^{eff}(t)$, respectively.  The magnetization vector ${\bf M}(r,t)$ in \Eq{MeffInt(r)} resulting from the applied rf is calculated starting from thermal equilibrium ${\bf M}_0=(0,0,1)$ using the rf amplitude $\nu_{rf}(r)$ given in \Eq{nu_rf(r)}.  

In the numerical simulations, the integration in \Eq{MeffInt(r)} is approximated by a discrete sum.  For each offset, the gradient (which depends on $\mathbf{M}$) is averaged over the range of rf spatial variation by one of the two methods outlined in Appendix A.  The first method samples $\mathbf{M}(r,t)$ at equally spaced $r$.  Since $\nu_{rf}$ varies as $1/r$, this has the effect of coarsely digitizing the rf for small values of $r$ and sets the stepsize $\Delta r$ required for accurate simulations of rf inhomogeneity in the toroid.  At large $r$, however, $\Delta r$ is more accurate than necessary.  Alternatively, the second method samples $\mathbf{M}[\nu_{rf}(r),t]$ at equally spaced $\nu_{rf}$, multiplied by the proper weight for each rf frequency that accurately represents the nonlinearity of $\nu_{rf}(r)$, as derived in Appendix A.
The gradients resulting from either procedure, averaged over the range of rf spatial variation for each individual offset, are subsequently averaged with equal weight over the range of offsets $\nu_{off}$ to give the overall gradient for the performance factor $\Phi$.

Optimizations for a range of experimental parameters obtained efficient and robust pulses for ratios $r_{max}/r_{min}$ as high as 100, which is significant for extending the present results to additional toroid applications.
An experimental upper limit $\nu_{rf}^{max} = 25$~kHz on $\nu_{rf}(r_{max})$ for our toroid cavity resonator ($r_{min}=1$~mm, $r_{max}=6$~mm) limited the time-dependent pulse amplitude at $r_{max}$ to the range $0\leq \nu_{rf}(r_{max},t) \leq \nu_{rf}^{max}$ in the optimizations.
We consider offsets $\nu_{\it off}$ in the range $\pm 1.5$ kHz, corresponding to a 15 ppm $^1$H chemical shift at a spectrometer frequency of 200 MHz.
Different experimental settings are accommodated by expressing frequencies relative to the limit $\nu_{rf}^{max}$. In these relative frequency units, 
the current experimental setting corresponds to an offset range of 
$-0.06 \leq \nu_{off}/\nu_{rf}^{max} \leq 0.06$.  Pulses were optimized starting from a set of random initial pulses. Each pulse was digitized in time steps of 0.5~$\mu$s duration and, at each time step $t_j$, the rf amplitude modulation function $a(t_j)$ and the phase $\phi(t_j)$ were optimized.

\subsection{Point-to-point (PP) pulses}

Point-to-point transformations rotate one specified initial state about a fixed axis to a desired target state.  For example, a $90_y^\circ$ PP pulse for the rotation $M_z \rightarrow M_x$ will not in general rotate any other component $90^\circ$ about the $y$-axis.  An optimal hard pulse achieves a maximum fidelity for this PP transformation of only 74\% \cite{Toroid_Review} (see Appendix B).  A conservative lower bound for the required duration of a composite or shaped PP excitation pulse to achieve excellent overall performance can be estimated in comparison to the duration of a 90$^\circ$ hard pulse with amplitude set to the toroid upper limit $\nu_{rf}^{max} = 25$~kHz available at $r_{max}$.  This results in a conservative estimate for the lower bound of $1/(4\,\nu_{rf}^{max})=  10$~$\mu$s. 
The minimum time to reach a fidelity of, e.g., 99\% is expected to be significantly longer than 10~$\mu$s, since this hard pulse would generate a 540$^\circ$ rotation at $r_{min}$.  Hence, additional time is needed to create uniform excitation for all rf amplitudes in the sample and all offsets of interest.

\begin{figure}[htbp]
\centering{
\begin{overpic}
[width=\linewidth]{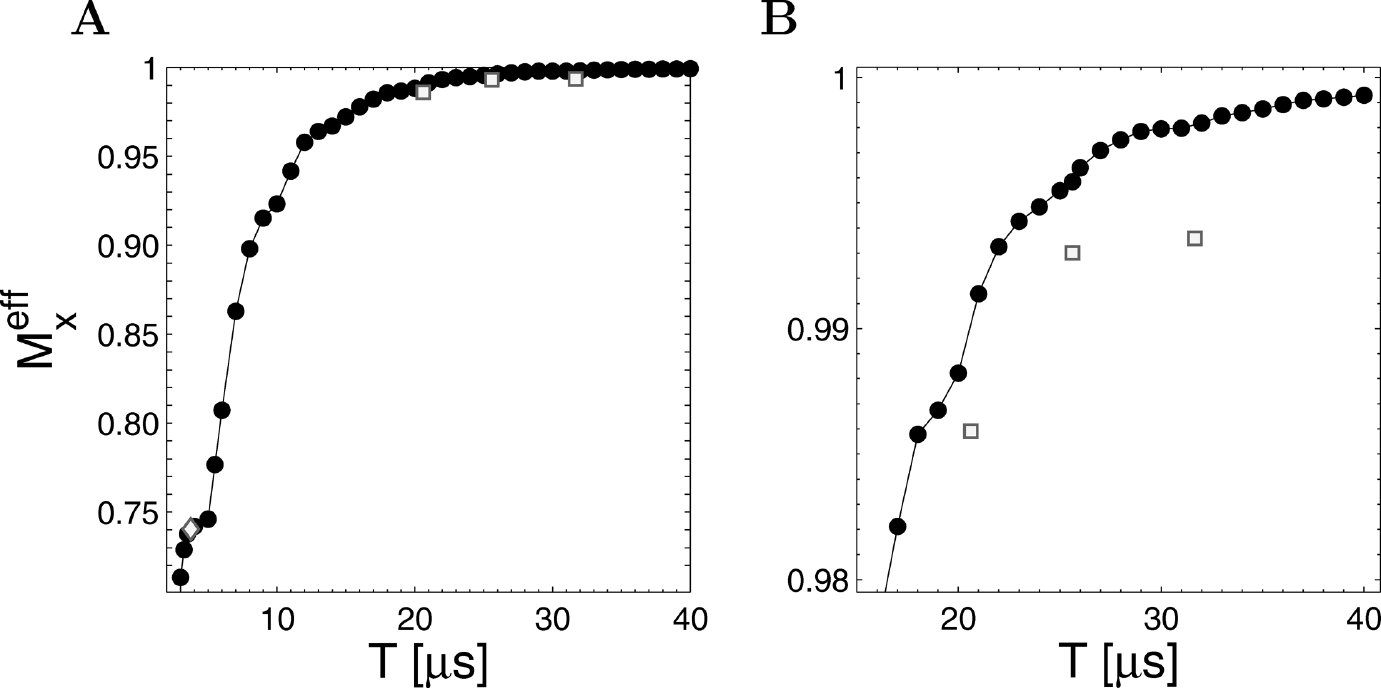}
\end{overpic}
}
\caption[TOP]
{
\label{TOP} \footnotesize \baselineskip 12pt Time-optimal pulse (TOP) curves showing optimal pulse performance as a function of pulse length for the PP transformation $M_z \rightarrow M_x$ applied to the toroid geometry and resonance offset range described in the text. The optimal length on resonance hard pulse derived in Appendix B ($T = 3.74\ \mu$s) is plotted (open diamond) for comparison.  $M^{eff}_x$ generated by three optimized composite pulses based on the phase-modulated excitation pulses in \cite{Woelk:1995} is shown as gray squares.  An enlarged view of the region where $M^{eff}_x \ge 0.98$ is shown in {\bf B}. Optimized composite pulse parameters are provided in the Supplementary Material.}
\end{figure}

To characterize the performance limits of robust excitation pulses, we considered pulse lengths $T$ in the range 2.5--40 $\mu$s.  Multiple optimizations were performed starting with random pulse shapes at each $T$ to obtain the best overall efficiency for transforming magnetization $M_z \rightarrow M_x$ in the toroid probe.  The resulting time-optimal pulse (TOP) curve is shown in \Fig{TOP}.
The TOP curve reaches a value of 99\% for the quality factor 
   \begin{equation}
\Phi_{PP} = M^{eff}_x
\label{PPQF}
   \end{equation}
at $T \approx 22\ \mu$s, which is only $2.2$ times longer than the conservative  estimate of 10 $\mu$s for the lower bound given above.

Existing composite pulse schemes \cite{Woelk:1995} that are also robust to a broad range of rf inhomogeneity were optimized for use in the toroid and are plotted (open squares) for comparison.  Pulse parameters are provided in the Supplementary Material.
For this experimental setting, a composite pulse constructed with an optimal duration of $25.62\ \mu$s achieves a quality factor of $99.30\%$. 
This pulse approaches the performance shown in the TOP curve of Fig. \ref{TOP}, indicating that it is close to the performance limit for this pulse duration.  However, a shorter optimized pulse of duration $ 22\ \mu$s is able to achieve the same quality factor as the composite pulse. 

\begin{figure}[htbp]
\centering{
\begin{overpic}
[width=\linewidth]{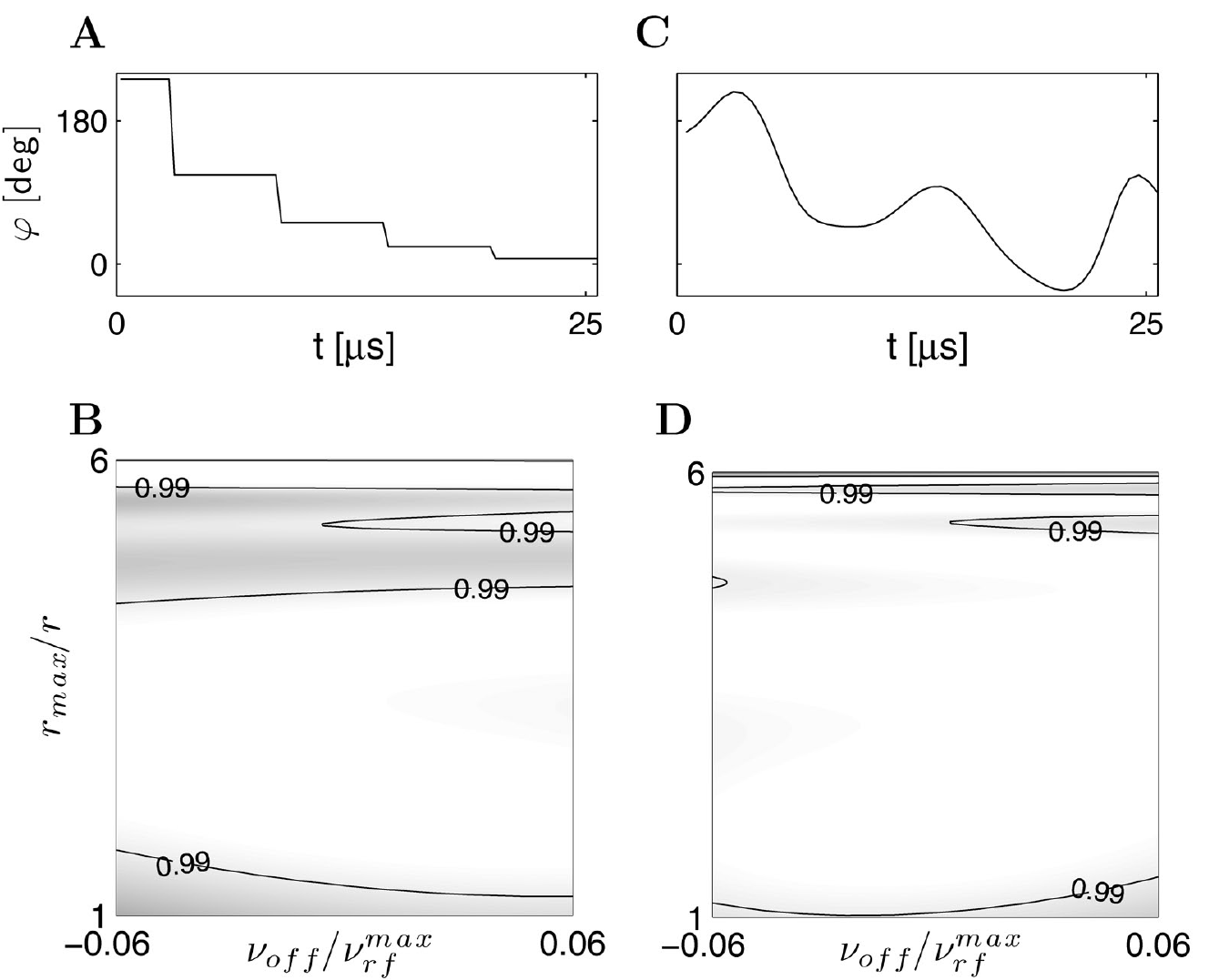}
\end{overpic}
}
\caption[9beta]
{
\label{composite_pulse} \footnotesize \baselineskip 12pt Comparison of the composite excitation pulse from Fig. \ref{TOP} with $T = 25.62\ \mu$s designed according to Ref.~\cite{Woelk:1995} and the optimzed pulse of equal length. Both pulses are purely phase-modulated and transform $M_z \rightarrow M_x$. \textbf{(A)} Composite pulse phase. \textbf{(B)} Excited magnetization plotted as a function of resonance offset $\nu_{off}$ and radial variation of the rf field strength $\nu_{rf}(r)$ given by \Eq{nu_rf(r)}.  Both axes are normalized to the maximum rf field $\nu_{rf}^{max} = 25$~kHz available at $r_{max}$ in the toroid probe. \textbf{(C)} and \textbf{(D)} Corresponding plots for the optimized pulse.  The composite pulse performance is very close to the optimized pulse, but, as shown in \Fig{TOP}, a shorter optimized pulse of length 22 $\mu$s can be used to achieve performance similar to the composite pulse.  }
\end{figure}

More detailed performance of this constant amplitude, phase-modulated composite pulse is shown in \Fig{composite_pulse} compared to an optimized phase-modulated pulse of the same length.  There are many different optimized pulses that achieve performance  similar to the pulse shown. This partcular pulse was chosen because the profile of its phase modulation is reminiscent of the composite pulse.  Pulses with even more striking similarity to the composite pulse shape can be found at $T = 29\ \mu$s with slightly improved performance.

\begin{figure}[htbp]
\centering{
\begin{overpic}
[width=.49\linewidth]{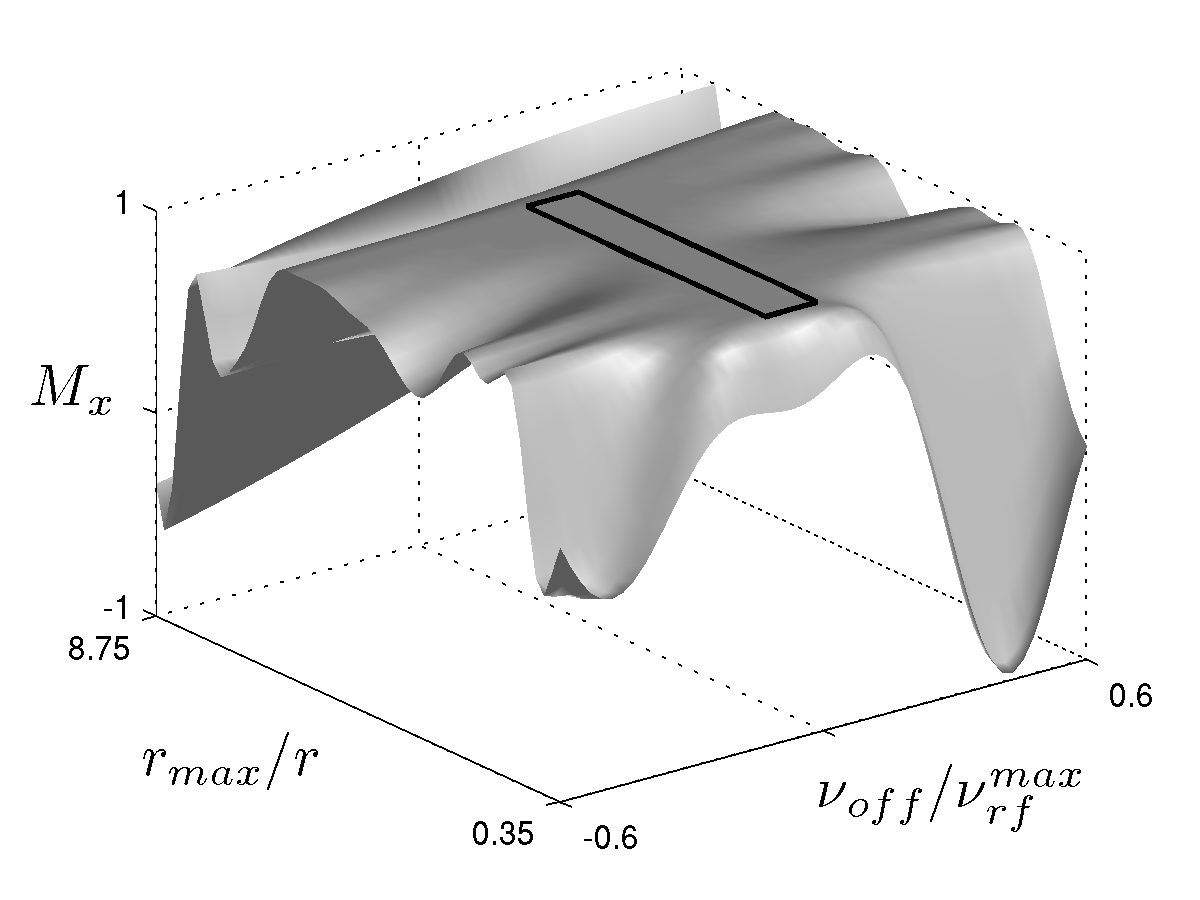}\put(42,75){\large {\sf simulation}} 
\end{overpic}
\begin{overpic}
[width=.49\linewidth]{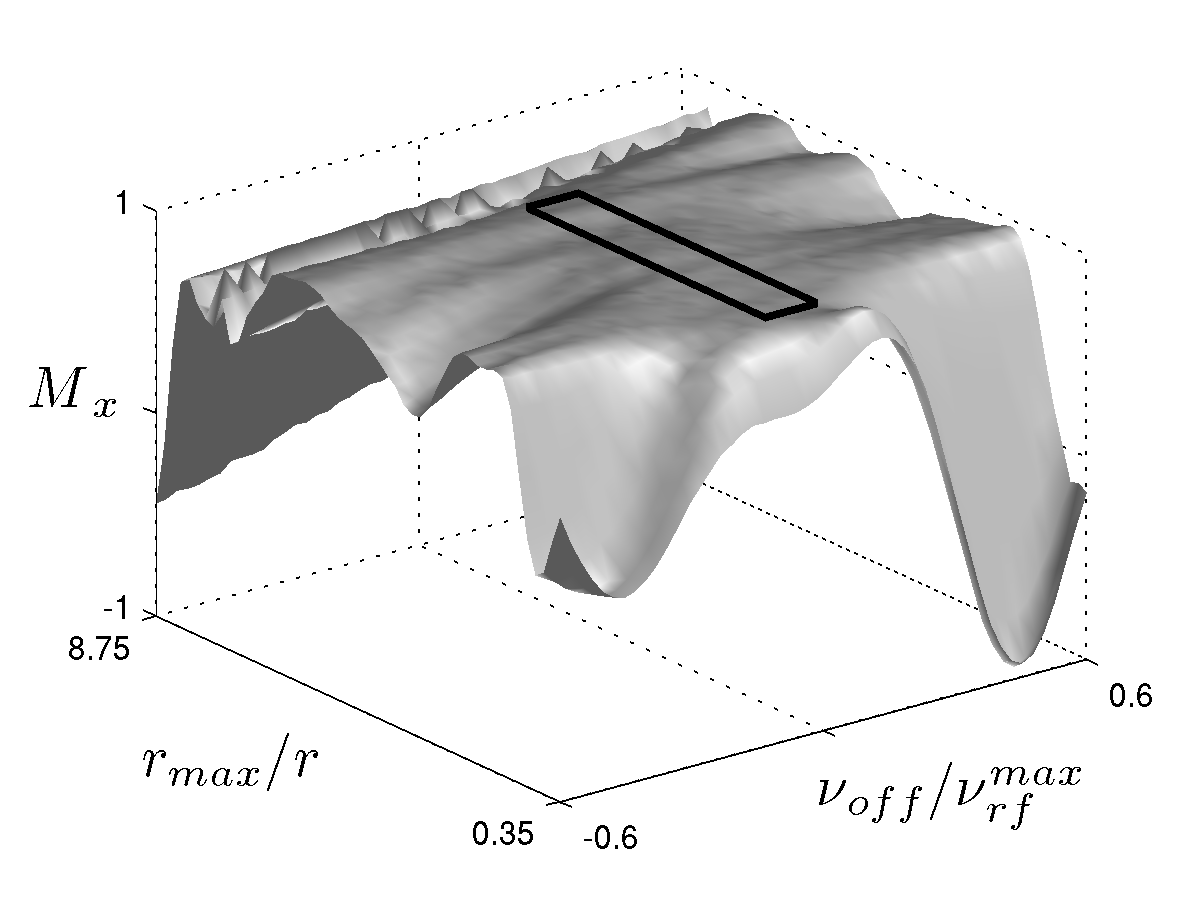}\put(40,74){\large {\sf experiment}} %
\end{overpic}
}
\caption[]
{
\label{naums_pulse_result} \footnotesize \baselineskip 12pt Simulated and experimental performance of a high-quality PP pulse ($T=100\ \mu s$) transforming $M_z \rightarrow M_x$ plotted as a function of resonance offset and radial variation of the rf field strength, as described in \Fig{composite_pulse}. The value of $M_x$ in the region for which the pulses were optimized (indicated by a black rectangle) is 0.999.}
\end{figure}

The close match that is possible to achieve between theoretical and experimental pulse performance is illustrated in \Fig{naums_pulse_result} for a high-fidelity excitation pulse ($T = 100\ \mu$s) that transforms $M_z \rightarrow M_x$ with a quality factor of 0.9993. The value of $M_x$ is plotted as a function of resonance offset and radial variation of the rf field strength, scaled as described in \Fig{composite_pulse}.  The region for which the pulse was optimized ($-0.06\leq \nu_{off}/\nu_{rf}^{max} \leq 0.06$ and $1 \leq \nu_{rf}(r)/\nu_{rf}^{max}  \leq 6$ ) is indicated by a box in the figure. 
The experiments were performed on a Bruker AV 250 spectrometer equipped with a conventional 5mm QNP probe. A Shigemi tube was filled to a height of 4 mm with $\approx 1 \%$ H$_2$O in D$_2$O doped with copper sulfate to enhance relaxation and reduce experimental time. In this experimental setting, the rf inhomogeneity is on the order of 1\% and can be neglected. We systematically scanned offset (50 steps) and the rf (50 steps) in a series of 2500 experiments.  An excellent match was found between the simulated and experimental performance of the pulse. 

The TOP curve thus provides a useful benchmark for past and future PP excitation pulses to be used in toroid probes.  Although improved PP excitation pulses were found in the current setting for $^1$H  applications of toroid probes, the improvement in excitation performance compared to previously developed composite pulses was relatively small. However, this improvement provides the option for decreasing pulse lengths by 10--15\%.  More importantly, the methods presented here can be applied to applications where larger offset ranges need to be covered, e.g., for $^{13}$C spectroscopy or higher spectrometer frequencies, where even greater performance gains compared to existing PP pulses can be expected.

\subsection{Universal rotation (UR) pulses}

Universal rotation pulses that transform any orientation of the initial magnetization about the same axis are desirable whenever more than one magnetization component has to be rotated in a controlled way.  This is the case for most multi-dimensional NMR experiments, which cannot be efficiently implemented based on PP pulses alone.  In the following, we characterize the performance of UR pulses suitable for use in the toroid probe, analogous to our treatment of PP pulses.  The primary difference is that our quality factor for assessing performance becomes \cite{GRAPE} 
   \begin{equation}
\Phi_{UR} = \Re \langle U_{target}\mid U_{eff}(T)\rangle,
\label{URQF}
   \end{equation}
where $U_{target}$ is the desired (ideal) propagator for the transformation, $U_{eff}(T)$ is the actual effective propagator at the end of the pulse, and the trace operator Tr sums the diagonal elements of the resulting matrix product. This is equivalent to combining (averaging) the quality factors for separate PP transformations about the coordinate axes.  For example, a 90$^\circ$ universal rotation of single-spin magnetization about the y-axis axis averages $\Phi_{PP}$ for $M_z \rightarrow M_x$, $M_x \rightarrow -M_z$, and $M_y \rightarrow M_y$.  

\begin{figure}[htbp]
\centering{
\begin{overpic}
[width=.49\linewidth]{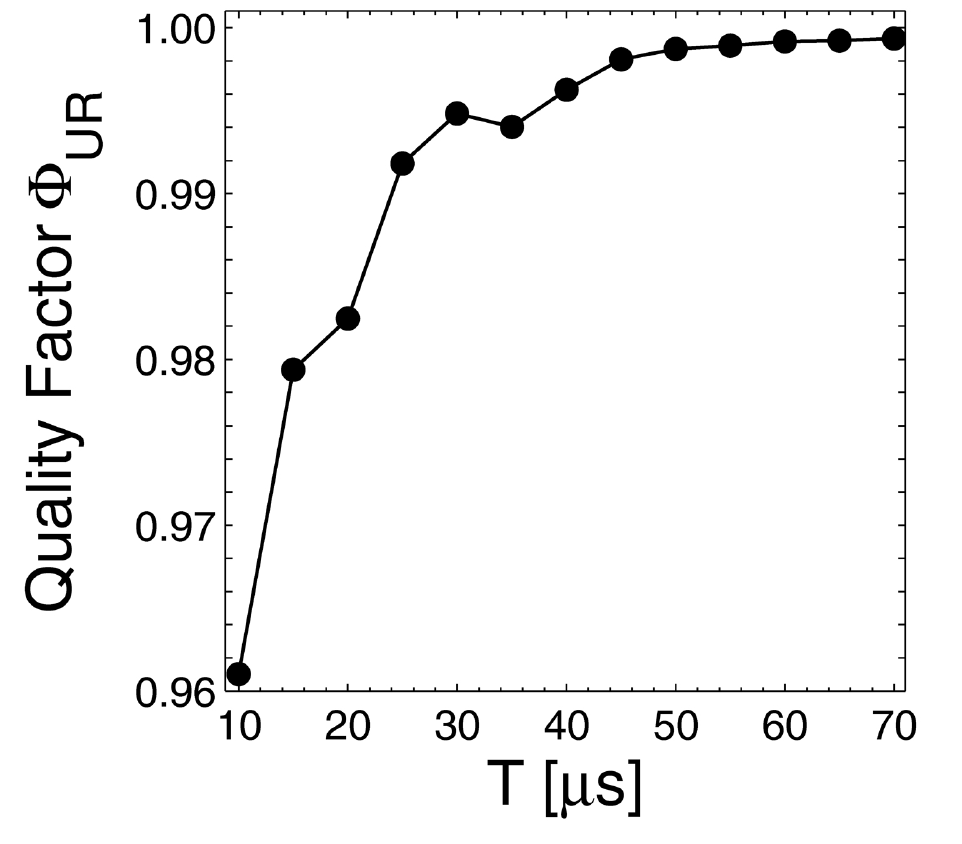}\put(7,90){\large {\bf A}} 
\end{overpic}
\begin{overpic}
[width=.49\linewidth]{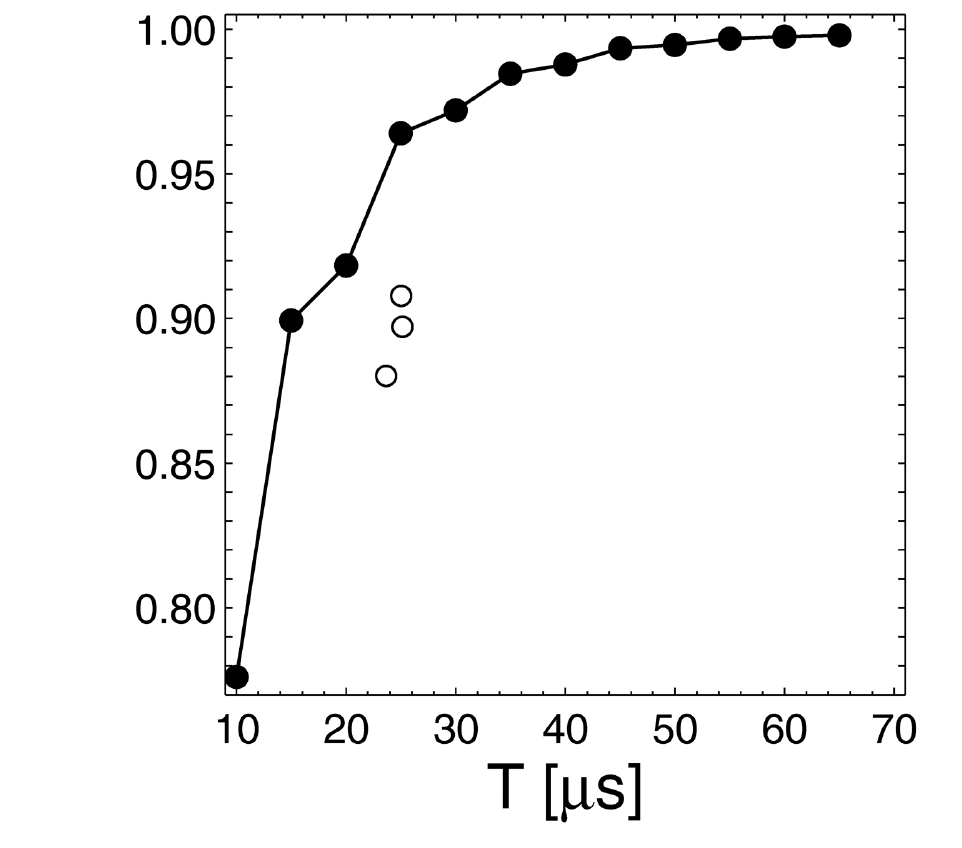}\put(7,90){\large {\bf B}} 
\end{overpic}
}
\caption{\footnotesize \baselineskip 12pt Time-optimal pulse (TOP) curves showing  optimal performance as a function of pulse length for the universal rotations \hbox{UR 90$^\circ_x$} ({\bf A}) and \hbox{UR 180$^\circ_x$} ({\bf B}) applied to the toroid geometry and resonance offset range described in the text. By comparison, the quality factor for the best optimally parameterized \cite{Skinner10} OP-BIR4 pulse we were able to generate for $T=50\ \mu s$ is 0.9588. The performance of three optimized UR 180$^\circ_x$ pulses constructed by the procedure of Ref.\cite{construction} from phase-modulated excitation pulses in \cite{Woelk:1995} is indicated by circles in ({\bf B}).  Optimized composite pulse parameters are provided in the Supplementary Material.}
\label{TOP_UR}
\end{figure}

\Figure{TOP_UR} shows the TOP curves we found for the performance limits of UR 90$^\circ$ and UR 180$^\circ$ pulses suitable for use in the toroid.  These pulses require approximately twice the time of PP pulses to approximate an ideal quality factor of one.  This is consistent with earlier work on the construction of UR pulses from two PP pulses \cite{Luy:2005}.  
The composite pulses of Ref.\cite{Woelk:1995} are not able to achieve the UR performance of a simple hard pulse in the toroid.  

To date, the only other UR pulses with sufficient tolerance to rf scaling and resonance offset that might make them suitable for use in toroid probes are the adiabatic BIR-4 pulses \cite{BIR4:1991}.  However, pulse lengths of $\sim 50\ \mu$s for the UR pulses obtained here, requiring an extremely rapid 12.5 $\mu$s half-passage in each BIR-4 segment, are too short for good adiabatic performance.  The peak performance we were able to obtain from BIR-4 at one particular combination of offset and rf scale factor was $\Phi_{UR} = 0.75$.  The performance over the volume of the toroid is considerably less.  The BIR-4 pulse was constructed using the tanh/tan functions for amplitude/frequency modulation \cite{BIR4:1991}.  For a given performance level, they allow shorter pulse lengths than other adiabatic modulation schemes \cite{Bendall:2002}.  Standard values of $\tan\kappa = 20$ and $\xi = 10$ or $\xi = 20$ were used for the tanh/tan shape parameters.  The frequency sweep was then adjusted for optimal performance.

Thus, standard BIR-4 is unsuitable for such short pulse lengths.  However, to more fully understand what is possible at $T = 50\ \mu$s, $\nu_{rf}^{max}$ = 25~kHz, and the desired range of resonance offset and rf tolerance, we optimized the shape parameters and the frequency sweep for the tanh/tan pulse using the OP algorithm  \cite{Skinner10} to obtain $\tan\kappa = 17$, $\xi = 71$, and a frequency sweep of 602.5~kHz for each half-passage.  The resulting quality factor, $\Phi_{UR} = 0.959$ over the full volume of the toroid volume, is at the lower limit of the acceptable performance range considered in \Fig{TOP_UR}.  However, this optimally parameterized OP-BIR4 pulse is not adiabatic and only retains the basic shape and form of \hbox{BIR-4}.  More dramatic deviations from adiabaticity were found for the parameter values $\kappa = 10.35$, $\xi = 20.5$, and frequency sweep 202.5~kHz, which produce the same $\Phi_{UR} = 0.958$.  This large value of $\kappa$, slightly greater than $6\,(\pi/2)$, produces 1.5 ``copies'' of the standard \hbox{BIR-4} frequency modulation scheme ($\kappa_{max} = \pi/2$) in each half-passage segment, while the tanh amplitude modulation still has one cycle varying between 0 and 1 in each segment.  The limits of non-adiabatic performance using standard amplitude/frequency modulation functions is beyond the scope of the present paper.  We simply note that a pulse length of 200 $\mu$s is needed before OP-BIR4 can achieve the same quality factor ($\Phi_{UR} = 0.999$) as the 50 $\mu$s pulses derived for the TOP curve.
\begin{figure}[htbp]
\vspace{-1cm}
\centering{
\begin{overpic}
[width=.9\linewidth]{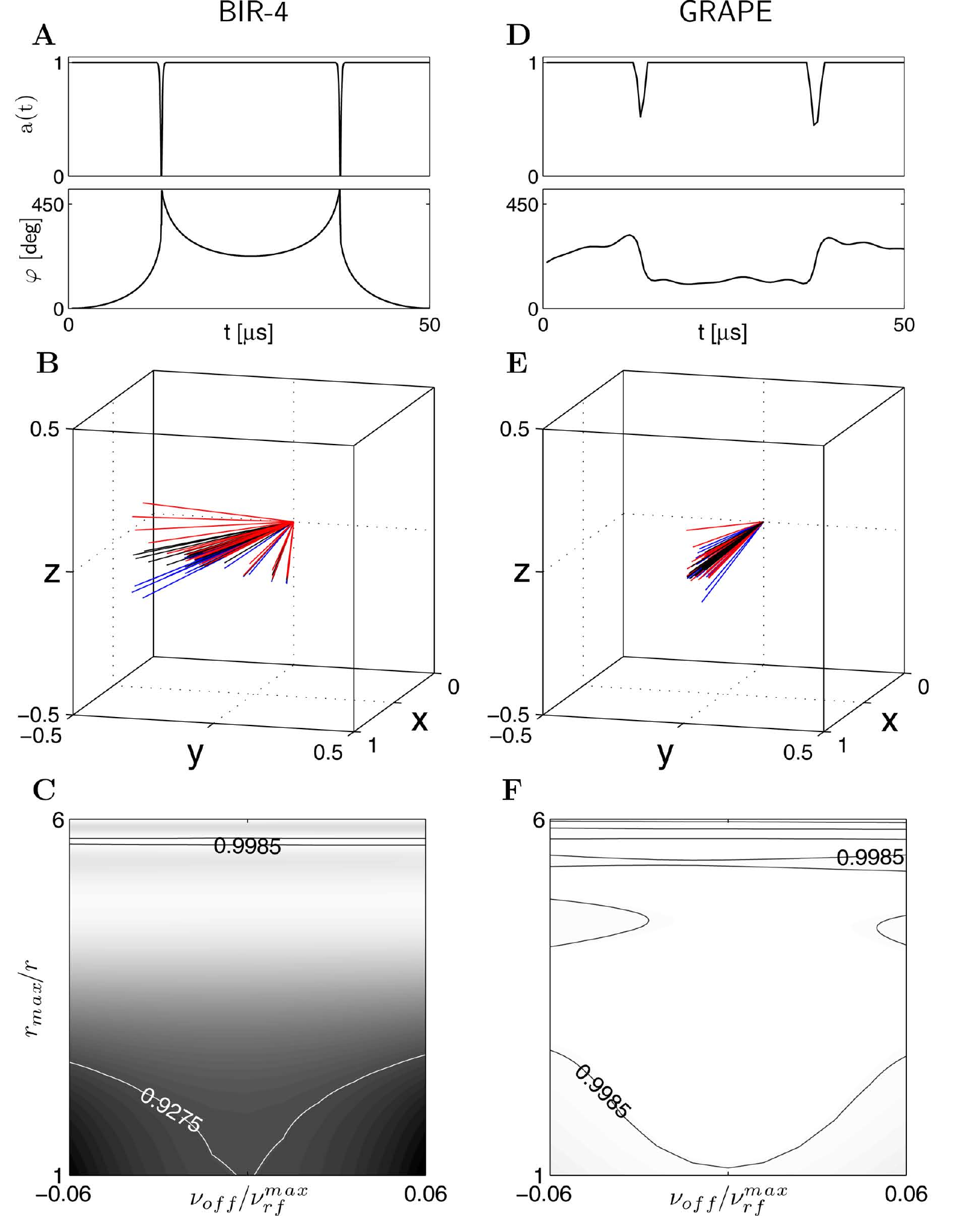}
\end{overpic}
}
\caption{\footnotesize \baselineskip 12pt The amplitude $a(t)$ and phase $\varphi(t)$ modulations of an optimally parameterized \cite{Skinner10} OP-BIR4 pulse (see text) are shown in ({\bf A}) compared to  one particular $90^{\circ}_y$ UR pulse of length 50 $\mu$s in ({\bf D}) that achieves the performance limit represented in the TOP curve of \Fig{TOP_UR}.  Neither optimized pulse is adiabatic, but the comparison illustrates the qualitative similarity the BIR-4 modulation scheme might have to a more generally optimized pulse.  The effective rotation axis is plotted for various offset and rf scale factors below each pulse, {\bf (B)} and {\bf (E)}, and the corresponding quality factor $\Phi_{UR} = \Re \langle U_{90^{\circ}_y}\mid U_{eff}(T)\rangle$ is plotted as a function of offset and rf scale factor in {\bf (C)} and {\bf (F)}.
The effective rotation axis for an ideal $90^{\circ}_y$ UR pulse is the $y$-axis for all offsets and rf scales.  For each (normalized) resonance offset $\nu_{off} / \nu_{rf}^{max}$ equal to $-0.06$ (red), $0$ (black) and $0.06$ (blue), 21 effective axes with rf scale factor $r_{max} / r$ ranging from 1 to 6 were calculated.  The effective rotation axes for pulses A and D show their deviation from the ideal, with further detail provided in the contour plots showing the quality factor achieved.  In contrast to these two optimized pulses, BIR-4 constructed using standard parameter values (see text) gives $\Phi_{UR} < 0.75$ over the desired range of offsets and rf scale factors.}
\label{90deg50us} 
\end{figure}

One particular $50\ \mu$s 90$^{\circ}_y$ UR pulse from the TOP curve set is shown in \Fig{90deg50us} along with details of its performance throughout the optimization  region.  This pulse was chosen for illustration because it has interesting qualitative similarities to the BIR-4 amplitude/phase modulation. However, the frequency modulation profile of the optimized pulse (obtained as the derivative of the phase modulation, but not shown) makes it clear how different it is from a standard adiabatic frequency sweep. \Figure{90deg50us} serves to illustrate the qualitative similarities that BIR-4 may have to an optimized pulse rather than the other way around. Adiabatic pulses are an elegant solution to the problem of large rf inhomogeneity, providing a simple physical picture that is easy to understand.  But optimal control shows there are many other solutions that are not so intuitive.

\subsection{Toroid NMR:  Implementation of DQF COSY}
\label{cosy}

To demonstrate the performance of the optimized pulses in two-dimensional experiments, we chose the DQF COSY experiment \cite{Rance:1983} as a particularly simple illustrative example.  Only the second and third pulses of this three-pulse sequence need to be UR pulses, but the same UR 90$^\circ$ pulse was used for all three pulses as a more stringent test of its capabilities. 

Two sets of experiments were performed using a conventional probe for the first set and a toroid probe of dimensions $r_{min}=1$ mm and $r_{max}=6$ for the second.  In each experiment, spectra were acquired using either conventional hard pulses for all pulses in the sequence or optimized pulses.

The first set of DQF COSY experiments were recorded on a Bruker AV 600 spectrometer at 600 MHz proton resonance frequency equipped with a conventional probe with negligible rf inhomogeneity. As a simple model system we used cytosine as a two spin system. The sample was prepared using a saturated solution of cytosine in a mixture of  DMSO-d$_6$/D$_2$O (10/1, vol/vol) which was doped \cite{Gutowsky:1963} with paramagnetic Chromium(III)acetylacetonate to reduce the longitudinal relaxation time  from  $3.7$ seconds, initially, to $0.15$ seconds to accelerate experimental data acquisition. 

The upper limit for the rf amplitude in the conventional probe is 30~kHz, which scales to 5~kHz to match the weakest rf amplitude in the toroid probe at $r_{max} = 6$~mm.  
Since this is one fifth the value the pulse was originally optimized for, $T$ must be increased by the same factor from  50 $\mu$s, originally, to 250 $\mu$s.  The carrier frequency was offset $300$~Hz from the \hbox{N-C\textit{H}=CH} diagonal peak, corresponding to the maximum (positive) optimized offset range $\nu_{off}/\nu_{rf}^{max} = \pm 0.06$.
The rf amplitude of the 90$^{\circ}$ hard pulse ($\nu_{rf} = 17.5 \mbox{kHz}, T=14.29\ \mu$s) was chosen at the middle of the range of rf variation (5--30~kHz), corresponding to an rf scale factor of $3.5$ times the minimum amplitude. 
\begin{figure}[htbp]
\vspace{1.5cm}
\centering{
\begin{overpic}
[width=\linewidth]{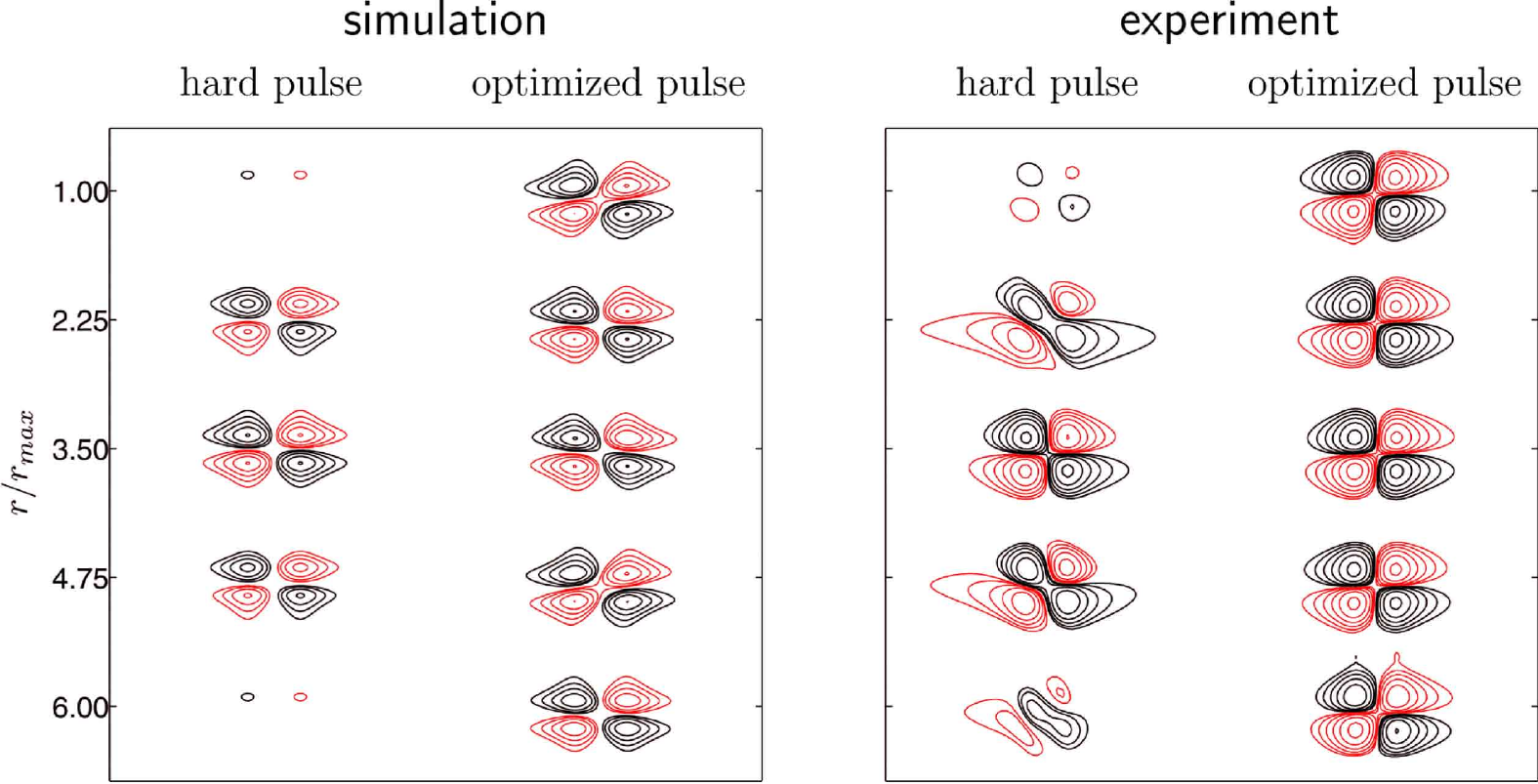}
\end{overpic}
}
\caption [Parts of DQF COSY spectra of cytosine]
{\footnotesize \baselineskip 12pt
Simulated (left) and experimental (right) DQF COSY spectra of cytosine.
The diagonal peak of N-C\textit{H}=CH is shown. 
Two series of spectra using rectangular and shaped pulses are displayed for $r_{max}/r=$ 1.00, 2.25, 3.50, 4.75, 6.00 (c.f. text) }
\label{simcosy}
\end{figure}

Experimental results for the \hbox{N-C\textit{H}=CH} diagonal peak are plotted as contours in \Fig{simcosy} for rf scale factors $r_{max} / r$ equal to 1, 2.25, 3.5, 4.75 and 6. Simulated results are shown on the left.  There is an excellent match between the simulated and experimental performance of the pulse.  Although hard pulses work only within a very limited range of $r_{max} / r$, the optimized UR pulse provides excellent performance over the entire rf spatial variation appropriate for the toroid.

\begin{figure}[htbp]
\centering{
\centering{
\begin{overpic}
[width=\linewidth]{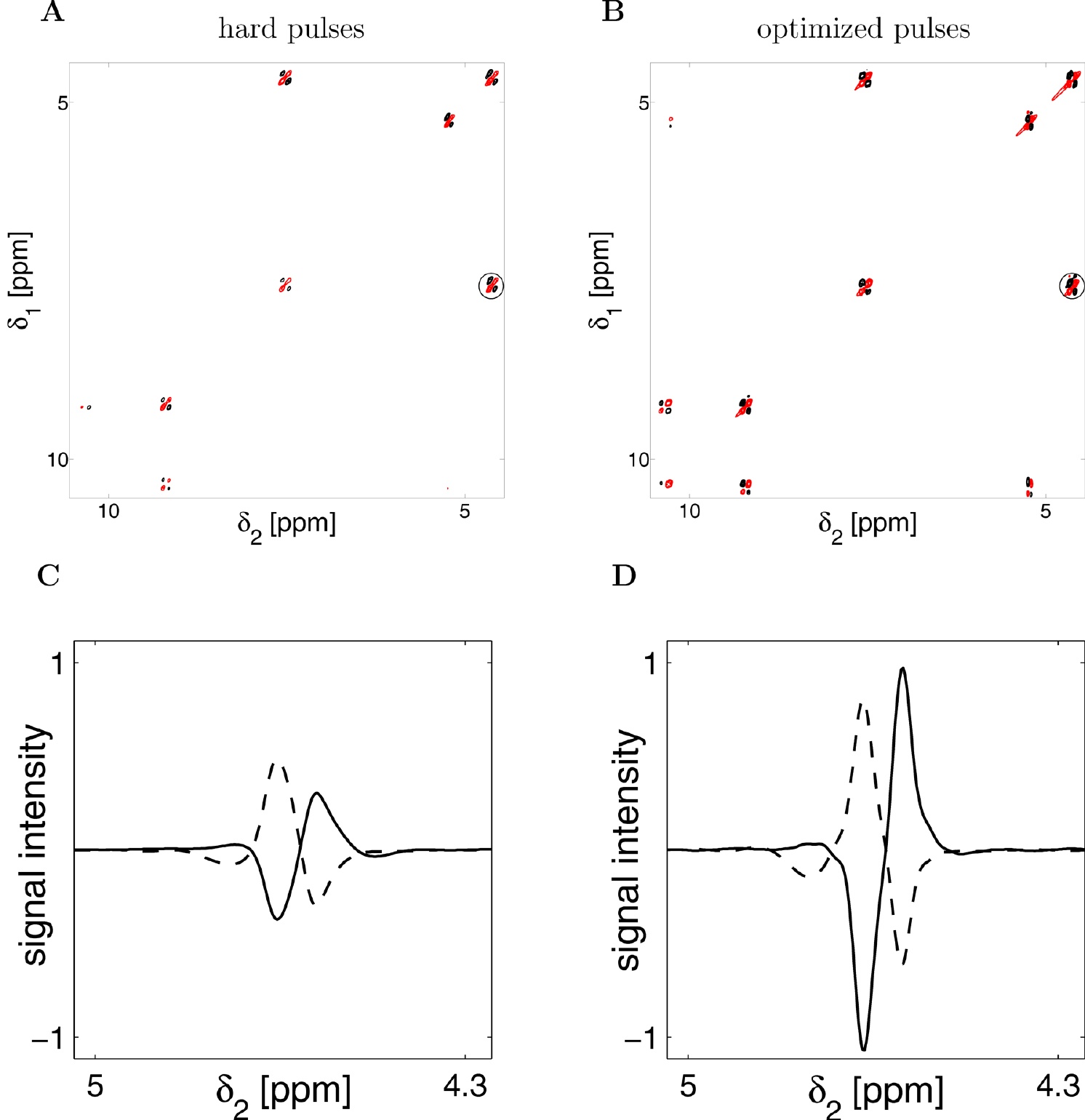}
\end{overpic}
}
}
\caption[toruscosy_vgl.png]
{\footnotesize \baselineskip 12pt
\label{toruscosy_vgl.png}DQF COSY spectra of ethyl crotonate recorded on a toroid probe using ({\bf A}) hard and ({\bf B}) optimized UR pulses. Slices through the cross peak circled in each spectrum are displayed in ({\bf C}) and ({\bf D}), showing a signal gain of 2.2 for the experiment using the optimized UR pulse.  Other peaks showed signal gains ranging from 1.7 to 2.7 relative to the hard pulse experiment.
}
\end{figure}

The second set of experiments was performed on a 200~MHz Bruker Avance DRX spectrometer using a homebuilt toroid probe with $r_{min}=1$~mm and $r_{max}=6$~ mm.  Details of the toroid design can be found in Ref.\cite{Trautner:2002a}.
Use of an earlier generation console produced systematic errors in the pulse waveform if the pulse generator did not have sufficient time to respond to changes in pulse phase.  To provide sufficient time, the pulse increment was first increased from 0.5 $\mu$s to 0.85 $\mu$s. Each pulse increment was preceded and followed by zero-amplitude sub-pulses of the same duration and phase to provide the necessary response time.  The net affect was to increase $T$ from 50$\mu$s to 357 $\mu$s.  This scales the optimized peak rf and resonance offset range of the original 50$\mu$s UR 90$^\circ$ pulse by the ratio of the pulse lengths.  Instead of adjusting experimental settings to fit the modified pulse, we optimized a new 90$^\circ$ UR pulse for the same rf and resonance offset parameters considered throughout the present paper, but with a duration of $T=357\ \mu$s, a pulse increment duration of 0.85 $\mu$s, and alternating zero-amplitude sub-pulses as described above.

Contour plots of complete 2D DFQ COSY spectra of ethyl crotonate obtained using a toroid probe are shown in \Fig{toruscosy_vgl.png} for rectangular (A) and optimized UR (B) pulses. The carrier frequency was set to 0 ppm for both spectra, so that the chemical shift range for which the pulse was optimized ranges from $-7.5$ ppm $\le \delta \le +7.5$ ppm.  Nevertheless, as the experimental results demonstrate, 
our UR pulse provides very good performance beyond the optimization range, up to +10~ppm.  Simulations indicate an extended range of $\pm10.5$ ppm in which the pulse provides acceptable performance.  Slices from the spectra are displayed in C and D, highlighting the cross peak marked by circles in \Fig{toruscosy_vgl.png} A and B. For this cross peak, a signal gain of 2.2 is observed relative to the hard pulse experiment. For the other peaks of the spectra, $1.7$ to $2.7$-fold signal gains were obtained.  Even larger gains can be expected for experiments with more pulses, such as the HMBC experiment.

\section{Conclusion}

We have demonstrated the power of optimal control based methods for deriving rf pulses that tolerate the large $B_1$ spatial inhomogeneity present in toroid probes.  We considered both point-to-point (PP) rotations that transform a particular initial state to a desired final state and universal rotations (UR) that transform any initial state about a well-defined fixed axis to a desired final state. The limits of PP and UR pulse performance as a function of pulse length were provided for a fairly typical toroid probe geometry that produces a factor of six variation in the $B_1$ field strength over the radial dimension of the toroid.  Such pulses are crucial for realizing the full potential of toroid detectors for unique applications such as in situ reaction studies at high-pressure and/or high temperature.  

Optimized 90$^\circ$ UR pulses were utilized in a DQF COSY experiment to implement the first two-dimensional spectroscopic application using a toroid probe.  Factors 1.7--2.7 gain in signal intensity were obtained compared to the conventional experiment using a hard pulse.  Even larger gains can be expected for other homonuclear and heteronuclear experiments that employ more pulses or require performance over a larger range of resonance offsets.  These kinds of experiments are important for future development of the field, but the present work is a significant step in opening the door to a wealth of new applications for toroid probes.

In addition to PP excitation pulses,  PP inversion pulses were also optimized and an example is given in the supplementary material.  Excitation and inversion pulses considered in the text are available in electronic form
at 

http://www.org.chemie.tu-muenchen.de/glaser/Downloads.html 

and can also be found in the supplementary material.

\vskip 2cm
\section{Acknowledgements}
This material is based upon work supported by the National Science Foundation under Grant Numbers CHE 0943441 to T.E.S. and CHE 0943442 to K.W.  M.B. thanks the Fonds der Chemischen Industrie for a Chemiefonds stipend. S.J.G. acknowledges support from the DFG (GI 203/6-1), SFB 631, the EU program Q-ESSENCE, the Fonds der Chemischen Industrie, and facilities support for experiments performed at the
Bavarian NMR center at TU Muenchen. 

\pagebreak
\section*{Appendix A: Discretizing methods for the calculation of the effective magnetization vector ${\bf M}^{eff}$}

The effective magnetization vector ${\bf M}^{eff}(t)$ is defined in Eq.\ (3) simply as an integral of the magnetization vector ${\bf M}(r,t)$ over the radius $r$ from ${r_{min}}$ to ${r_{max}}$.
Here we consider two different methods to approximate the integral by a finite sum.

\vskip 1em

{\bf Discretization method A based on equidistant steps $\Delta r\,$}: 

The integration for ${\bf M}^{eff}(t)$ in \Eq{MeffInt(r)} can be approximated by sampling the radius $r$ at $N$ equidistant values $r_k\  (k = 1,2,\ldots,N)$ separated by $\Delta r = (r_{max}-r_{min})/N$, with 
$r_1 = r_{min} + \Delta r/2$ and $r_N = r_{max} - \Delta r/2$, to give
       \begin{equation}
{\bf M}^{eff}(t)\approx \frac{1}{N} \sum_{k=1}^N \ {\bf M}(r_k,t).
\label{MeffSum(r)}
       \end{equation}

This uniform sampling in $r$, however, corresponds to a non-uniform sampling of the rf amplitude $\nu_{rf}(r_k)$ given by \Eq{nu_rf(r)}.
For $r_{min} = 1$~mm, $r_{max}=6$~mm, and $\nu_{rf}^{max}=25$~kHz, 
the rf frequency difference between $\nu_{rf}(1\ \mathrm{mm})=150$~kHz and
$\nu_{rf}(1.5\ \mathrm{mm})=100$~kHz is 50~kHz, whereas the rf frequency difference between $\nu_{rf}(5.5\ \mathrm{mm})=27.3$~kHz and  $\nu_{rf}(6\ \mathrm{mm})=25$~kHz is only 2.3 kHz. A fine digitization is required for accuracy at small $r$ that is then not necessary at large $r$.  Method A can, therefore, be somewhat inefficient.

{\bf Discretization method B based on equidistant steps $\Delta \nu_{rf}\,$}: 

Writing $r = r_{max}\, \nu_{rf}(r_{max})\, / \, \nu_{rf}(r)$ 
from \Eq{nu_rf(r)} gives 
${\rm d}r = -r_{max}\, \nu_{rf}(r_{max})\, / \,\nu_{rf}^2\, {\rm d}\nu_{rf}$, and
the integral over the radius $r$ in \Eq{MeffInt(r)} can be replaced by an integral over the rf amplitude $\nu_{rf}$:
       \begin{equation}
{\bf M}^{eff}(t)=    
\frac{\nu_{rf}(r_{max})\,\nu_{rf}(r_{min})} 
     {\nu_{rf}(r_{min}) - \nu_{rf}(r_{max})} 
\int_{\nu_{rf}(r_{max})}^{\nu_{rf}(r_{min})} \frac{1}{\nu_{rf}^2}\,
{\bf M}(\nu_{rf},t) {\rm d}\nu_{rf}.
\label{MeffInt(rf)}
       \end{equation}
${\bf M}^{eff}(t)$ can be approximated by sampling the rf amplitude $\nu_{rf}$ at $N$ equidistant values $\nu_{rf,k}\ (k = 1,2,\ldots,N)$ separated by $\Delta \nu_{rf} = [\,\nu_{rf}(r_{min}) - \nu_{rf}(r_{max})\,]\,/\,N$, with $\nu_{rf,1} = \nu_{rf}(r_{max}) + \Delta \nu_{rf}/2$ and $\nu_{rf,N} = \nu_{rf}(r_{min}) - \Delta \nu_{rf}/2$, to give 
       \begin{equation}
{\bf M}^{eff}(t)\approx \frac{1}{N} \sum_{k=1}^N \ g_k \  {\bf M}(\nu_{rf, k},t)
\label{MeffSum(rf)}
       \end{equation}
with the weighting factors 
       \begin{equation}
g_k=\frac{\nu_{rf}(r_{max})\, \nu_{rf}(r_{min})} 
         {\nu_{rf, k}^2}.
\label{MeffWeights}
       \end{equation}

In contrast to method A, method B samples the rf amplitude in equidistant steps, albeit with non-uniform weights for the different rf amplitudes. According to \Eq{MeffWeights}, the relative weight $g_k$ for sample $k$ is inversely proportional to the square of the rf amplitude. Hence, samples with low rf amplitude contribute most to the effective magnetization vector and to the detected signal. For the example considered in method A, with $\nu_{rf}(r_{max}) = 25$~kHz and $\nu_{rf}(r_{min}) = 150$~kHz, the weight $g(r_{max}) = \nu_{rf}(r_{min})/\nu_{rf}(r_{max})$ is equal to 6,
whereas $g(r_{min})=\nu_{rf}(r_{max})/\nu_{rf}(r_{min})$ is only 1/6.

\section*{Appendix B: Optimal rectangular on-resonance excitation pulse}

For the toroid probe employed in the experiments, $r_{min} = 1$~mm, $r_{max}=6$~mm, the limit for the maximum rf amplitude available at $r_{max}$ is 25~kHz, which increases to 150~kHz at $r_{min}$.  Under these conditions, what is the optimal hard pulse length $T_{opt}$ that maximizes $[M^{eff}(T)]_x$ in \Eq{MeffInt(r)}?
If we (incorrectly) choose a linear weighting of the rf amplitudes, for example, 
the duration of an ideal rectangular $90^\circ$ pulse with an rf amplitude corresponding to $[\,(\nu_{rf}(r_{max}) + \nu_{rf}(r_{min})\,]/2 = 87.5$~kHz would be 2.86~$\mu$s.  However, the non-linear weighting given in \Eq{MeffWeights} emphasizes the contribution of smaller rf amplitudes, giving a longer value for $T_{opt}$.

\begin{figure}[htbp]
\centering{
\begin{overpic}
[width=.49\linewidth]{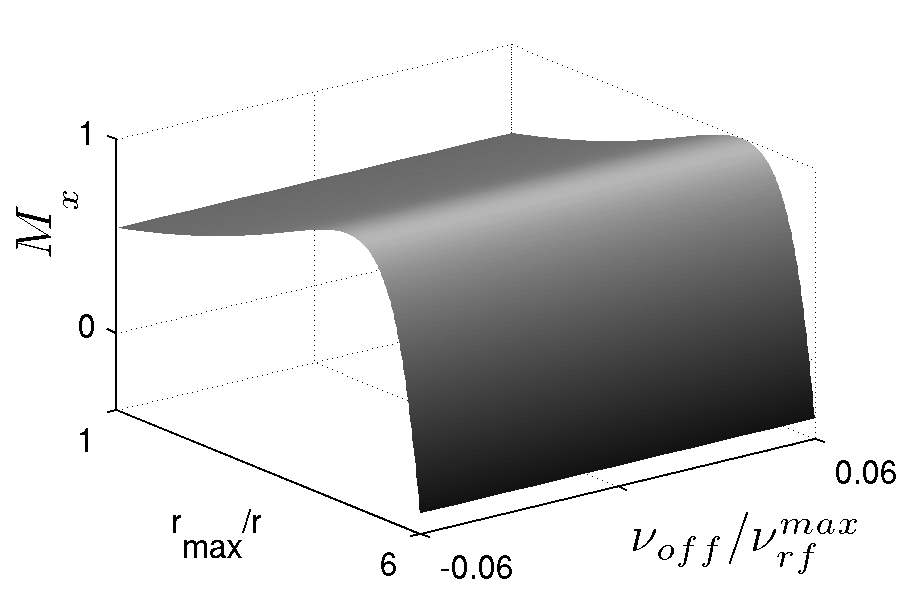}\put(2,60){\large {A}}
\end{overpic}
\begin{overpic}
[width=.49\linewidth]{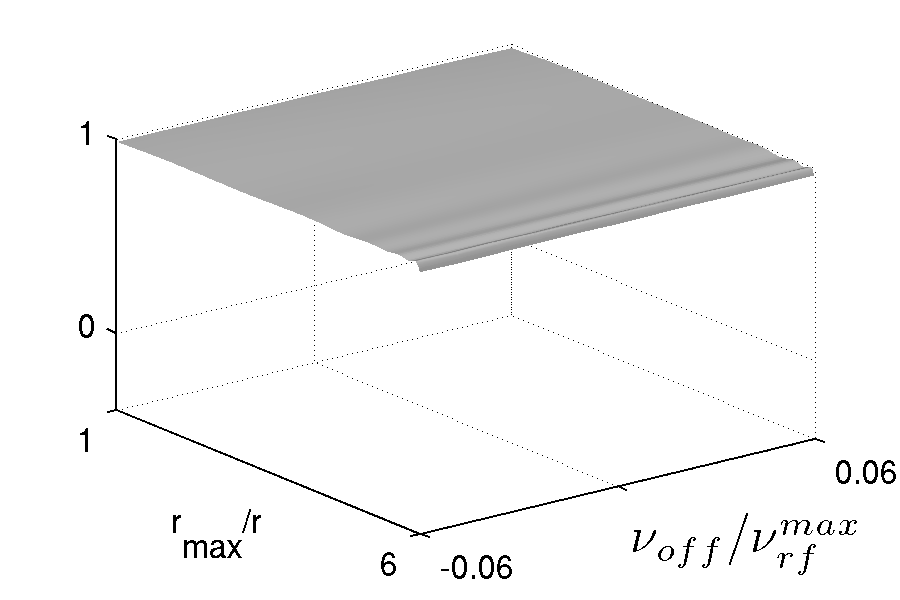}\put(2,60){\large {B}}
\end{overpic}
}
\caption[]
{
\label{comp_rect_OCT} \footnotesize \baselineskip 12pt {\bf (A)} The x-magnetization excited by a rectangular pulse of length $T_{opt}$ applied in the toroid probe is plotted as a function of (normalized) resonance offset $\nu_{off} / \nu_{rf}^{max}$ and rf scale factor $r_{max} / r$.  {\bf (B)}\ equivalent plot for an optimized pulse at $T=25.62\ \mu s$ from Fig \ref{TOP}.}
\end{figure}

A rectangular pulse of amplitude $\nu_{rf}$, duration $T$, and phase $y$ rotates the thermal equilibrium magnetization on resonance by an angle $2\pi\nu_{rf}\,T$ about the $y$-axis. 
The optimal length of an on-resonance rectangular pulse can therefore be determined numerically by substituting $M_x(\nu_{rf,k}\, ,t) = \sin(2\pi\nu_{rf,k}\,t)$ in \Eq{MeffSum(rf)}, with $\nu_{rf}(r_{max}) = 25$~kHz, and $\nu_{rf}(r_{min}) = 150$~kHz.
We find $T_{opt} = 3.74\ \mu$s.  The magnetization $M_x$ excited by this pulse is plotted in \Fig{comp_rect_OCT}.  It produces a signal that is 74\% of the signal obtained by an ideal $90^\circ$ pulse of the same length ($\nu_{rf} = 1/(4\,T_{opt}) = 66.8$~kHz) in the absence of any rf inhomogeneity.  The performance of an optimized pulse with $T = 25.62\ \mu$s, corresponding to the composite pulse of the same length in \Fig{TOP}, is plotted in 
\Fig{comp_rect_OCT}B for comparison.

\vskip 2cm
\section{Supplementary Material}

Pulse shapes for selected pulses from the text are given as Bruker pulse shape files in text format.

3.2\_CompositePPzx\_4PhaseSteps\\
3.2\_CompositePPzx\_5PhaseSteps\\
3.2\_CompositePPzx\_6PhaseSteps\\
Composite PP pulses designed according to \cite{Woelk:1995}. Cf. Fig. 1 and 2.

3.2\_OptimizedPPzx\_25point62us\\
Optimized PP pulses performing the transformation $M\_z \rightarrow M\_x$. Cf. Fig. 1 and 2.

3.3\_BIR-4\_50us\\
3.3\_OptimizedUR90\_50us\\
Optimized OP-BIR-4 and UR 90$^{\circ}\_y$-pulses. Cf. Fig. 5.

3.3\_OptimizedUR90\_70us\\
Optimized UR 90$^{\circ}_y$-pulses. Cf. Fig. 4 A.

3.3\_CompositeUR180\_9PhaseSteps\\
3.3\_CompositeUR180\_11PhaseSteps\\
3.3\_CompositeUR180\_13PhaseSteps\\
3.3\_OptimizedUR180\_65us\\
Composite and optimized UR 90$^{\circ}_y$-pulses. Cf. Fig. 4 B.

3.4\_OptimizedUR90\_357us\\
Optimized UR 90$^{\circ}_y$-pulse. This pulse was used to produce the spectra shown in Fig. 7 B and D.

{\baselineskip 16pt
3.5 Phase-modulated constant amplitude ($\nu_{rf}^{max}=25$~kHz) composite pulses based on the scheme of Ref.\cite{Woelk:1995}:
\squishlist
   \item[] PP 90$_x$
      \squishlist
         \item[] Four phase steps ($\phi = 225.1^{\circ}$, $\Delta t=2.946\ \mu$s)
            \squishlist
               \item[] $(\Delta t)_{\phi}(2\Delta t)_{\phi-120^{\circ}}
                      (2\Delta t)_{\phi-180^{\circ}}(2\Delta t)_{\phi-210^{\circ}}$
             \squishend
         \item[] Five phase steps ($\phi = 232^{\circ}$, $\Delta t=2.847\ \mu$s)
            \squishlist
               \item[] $(\Delta t)_{\phi}(2\Delta t)_{\phi-120^{\circ}}
                       (2\Delta t)_{\phi-180^{\circ}}
                       (2\Delta t)_{\phi-210^{\circ}}(2\Delta t)_{\phi-225^{\circ}}$
             \squishend
         \item[] Six phase steps ($\phi = 234.2^{\circ}$, $\Delta t=2.879\ \mu$s)
            \squishlist
               \item[] $(\Delta t)_{\phi}(2\Delta t)_{\phi-120^{\circ}}
                      (2\Delta t)_{\phi-180^{\circ}}
                      (2\Delta t)_{\phi-210^{\circ}}(2\Delta t)_{\phi-225^{\circ}}
                      (2\Delta t)_{\phi-232.5^{\circ}}$
             \squishend
        \squishend
   \item[] PP 180$_x$
      \squishlist
         \item[] Four phase steps ($\Delta t = 1.314\ \mu$s)
            \squishlist
               \item[] $(\Delta t)_{0^{\circ}}(2\Delta t)_{-120^{\circ}}
                       (2\Delta t)_{-180^{\circ}}(2\Delta t)_{-210^{\circ}}
                       (4\Delta t)_{-225^{\circ}}(2\Delta t)_{-210^{\circ}}
                       (2\Delta t)_{-180^{\circ}}(2\Delta t)_{-120^{\circ}}
                       (\Delta t)_{0^{\circ}}$
             \squishend
         \item[] Five phase steps ($\Delta t = 1.137\ \mu$s)
            \squishlist
               \item[] $(\Delta t)_{0^{\circ}}$\-$(2\Delta t)_{-120^{\circ}}$\-$
                       (2\Delta t)_{-180^{\circ}}$\-$(2\Delta t)_{-210^{\circ}}$\-$
                       (2\Delta t)_{-225^{\circ}}$\-$(4\Delta t)_{-232.5^{\circ}}$\-
                       $(2\Delta t)_{-225^{\circ}}$\-$(2\Delta t)_{-210^{\circ}}$\-
                       $(2\Delta t)_{-180^{\circ}}$\-$(2\Delta t)_{-120^{\circ}}$\-
                       $(\Delta t)_{0^{\circ}}$
             \squishend
         \item[] Six phase steps ($\Delta t=0.966\ \mu$s)
            \squishlist
               \item[] $(\Delta t)_{0^{\circ}}$\-$(2\Delta t)_{-120^{\circ}}$\-
                       $(2\Delta t)_{-180^{\circ}}$\-$(2\Delta t)_{-210^{\circ}}$\-
                       $(2\Delta t)_{-225^{\circ}}$\-$(2\Delta t)_{-232.5^{\circ}}$\-
                       $(4\Delta t)_{-236.25^{\circ}}$\-$(2\Delta t)_{-232.5^{\circ}}$
                       \-$(2\Delta t)_{-225^{\circ}}$\-$(2\Delta t)_{-210^{\circ}}$
                       \-$(2\Delta t)_{-180^{\circ}}$\-$(2\Delta t)_{-120^{\circ}}
                       (\Delta t)_{0^{\circ}}$
             \squishend
        \squishend
\squishend
}

\vskip 4em
\vfill \eject

\vfill \eject

\end{document}